\documentstyle[aps,epsfig]{revtex}

\begin{document}
\def\be{\begin{equation}}
\def\ee{\end{equation}}
\def\bfi{\begin{figure}}
\def\efi{\end{figure}}
\def\bea{\begin{eqnarray}}
\def\eea{\end{eqnarray}}
\newcommand{\ket}[1]{\vert#1\rangle}
 \newcommand{\bra}[1]{\langle#1\vert}
\def\id{\mathbb{I}}

\title{Effective temperature in the quench of coarsening systems to and below $T_C$}

\author{Federico Corberi$^\ddag$, Eugenio Lippiello$^\dag$ 
and Marco Zannetti$^\S$}
\address {Istituto Nazionale di Fisica della Materia, Unit\`a
di Salerno and Dipartimento di Fisica ``E.Caianiello'', 
Universit\`a di Salerno,
84081 Baronissi (Salerno), Italy}
\maketitle

\ddag corberi@sa.infn.it  
\dag lippiello@sa.infn.it
\S zannetti@na.infn.it

\begin{abstract}
We overview the general scaling behavior of the effective temperature in the quenches of simple non disordered
systems, like ferromagnets, to and below $T_C$. Emphasis is on the behavior as dimensionality is varied. Consequences
on the shape of the asymptotic parametric representation are derived. In particular, 
this is always trivial in the critical quenches with $T_C>0$. 
We clarify that the quench to $T_C=0$ at the lower critical dimensionality $d_L$, cannot be regarded as a critical quench. 
Implications for the behavior of the exponent $a$ of the aging response function in the quenches
below $T_C$ are developed.

\end{abstract}

\vspace{1cm}

{\bf keywords}

 Coarsening processes (Theory), Slow dynamics and aging (Theory)

\section{Introduction} \label{sec1}

In the study of slow relaxation phenomena~\cite{Cugliandolo2002}, typically, a temperature quench is made at time $t=0$ and 
the quantities
of interest, such as the autocorrelation $C(t,t_w)$ and the linear response function $R(t,t_w)$, are
monitored at subsequent times $0 < t_w < t$. The peculiar and interesting feature of these processes is that
equilibrium is never reached, namely there does not exist a finite time scale $t_{eq}$ such that for $t_w> t_{eq}$
two time quantites become time translation invariant.

A substantial progress in the study of these phenomena has been achieved after Cugliandolo and Kurchan~\cite{C-K} have introduced
the modification of the fluctuation-dissipation theorem (FDT)  as a quantitative indicator of the deviation 
from equilibrium~\cite{Crisanti}.
This can be formulated in several equivalent ways. Here, we find convenient to use the effective temperature~\cite{Peliti} 
\be
T_{eff}(t,t_w)=  {\partial_{t_w} C(t,t_w)  \over R(t,t_w) }
\label{0.1bis}
\ee
which coincides with the temperature $T$ of the thermal bath if the FDT holds, that is in equilibrium, while 
it is different from $T$ off equilibrium. Because of the absence of a finite equilibration time, the characterization of the system
staying out of equilibrium for an arbitrarily long time is contained in $\lim_{t_w \to \infty}T_{eff}(t,t_w)$. 
However, since $t_w \rightarrow \infty$ implies also $t \rightarrow \infty$, to make the limit
operation meaningful one must specify how the two times are pushed to infinity. One way of doing this is 
to fix the value of the autocorrelation function $C(t,t_w)$
as $t_w \rightarrow \infty$. More precisely, since $C(t,t_w)$ for a given $t_w$ is a monotonously decreasing
function of $t$, the time $t$ can be reparametrized in terms of $C$ obtaining
$T_{eff}(t,t_w) = \widehat{T}_{eff}(C,t_w)$ and, keeping $C$ fixed, the limit
\be
{\cal T}(C) = \lim_{t_w \to \infty} \widehat{T}_{eff}(C,t_w)
\label{2.15bis}
\ee
defines the effective temperature in the time sector corresponding to the chosen value of $C$. This quantity is important because, 
when appropriate conditions are satisfied,
provides the connection between dynamic and static properties~\cite{FMPP} through the relation
\be
P(q) = T {d \over dC} {\cal T}(C)^{-1}|_{C=q}
\label{fmpp}
\ee
where $P(q)$ is the overlap probability function 
of the equilibrium state. 
Therefore, different shapes of ${\cal T}(C)$
are associated to different degrees of complexity
of the equilibrium state.
A first broad classification of systems has been made~\cite{Cugliandolo99,Parisi99} 
drawing the distinction between coarsening systems, 
structural glasses and the infinite range spin glass model on the basis of the patterns displayed by  
${\cal T}(C)$, or related quantities such as the fluctuation-dissipation ratio (FDR) and the
zero field cooled (ZFC) susceptibility.
In this paper we focus on coarsening systems, exemplified by 
a non disordered and non frustrated system such as a ferromagnet, relaxing via domain growth after
a quench to or below the critical point~\cite{Bray,Godreche,Calabrese}. We show that even within
the framework of these so called trivial systems, the spectrum of the behaviors of ${\cal T}(C)$ 
is quite rich and interesting. 

In order to illustrate the problem, in Fig.1 we have drawn schematically the phase diagram,
in the temperature-dimensionality plane.
The critical temperature  vanishes at the lower critical dimensionality $d_L$. 
The disordered and ordered phase are, respectively, above and below the critical line $T_C(d)$.
Slow relaxation arises for quenches with the final temperature everywhere in the shaded area,
including the boundary. 
Usually, one considers some fixed $d>d_L$ (dashed line in Fig.1) and studies the quench to a 
final temperature $T \leq T_C$. The form of ${\cal T}(C)$ in the critical quench with $T=T_C$ (Fig.2a)
is different from that with $T<T_C$ (Fig.2b) and both are independent of $d$.
The interesting question, then, is what happens at the special point $(d_L,T=0)$, where critical
and subcritical quenches merge. Namely, will ${\cal T}(C)$ display continuity with the critical or with the
subcritical shape? On the basis of the available evidence from analytical 
solutions~\cite{Lippiello2000,Godreche2000,Corberi2002,Corberi2001}  and
numerical simulations~\cite{Generic} , the surprising answer is with neither of them. A {\it third} and nontrivial
form of ${\cal T}(C)$ is found (Fig.2c) in the quench to $(d_L,T=0)$. 

In this paper we present the scaling framework which allows to account in a unified way for this wealth
of behaviors displayed by ${\cal T}(C)$ upon letting $T$ and $d$ to vary. This, in turn, allows to gain 
insight on the challanging problem~\cite{Corberi2001,Corberi2003}  of the scaling of $R(t,t_w)$ in the quenches below $T_C$. Furthermore,
on the basis of these ideas, predictions can be made for the interesting case of the XY model. 
In that case, the phase diagram of Fig.1 is enriched by the presence of the Kosterlitz-Thouless (KT) line of
critical points at the lower critical dimensionality $(d_L=2)$, with $T \leq T_{KT}$. 
Therefore, slow relaxation in the XY model 
arises also by arranging the quenches onto the KT line. Since this is a critical line, the phenomenology
for $0 < T \leq T_{KT}$ is expected to be akin to the one along the $T_C(d)$ line, 
while at $T=0$ the switch to the third nontrivial form of ${\cal T}(C)$ is expected to take place. 

The paper is organised as follows. In section~\ref{sec2} the scaling properties of $C(t,t_w)$ and $R(t,t_w)$ are
reviewed. In section~\ref{sec3}  the general scaling behavior of $T_{eff}(t,t_w)$ is derived, and in section~\ref{sec4} the 
parametric form ${\cal T}(C)$ is obtained. The problem of the exponent $a$ of 
$R(t,t_w)$ below $T_C$ is presented in section~\ref{sec5}. Sections~\ref{sec6} and~\ref{sec7} are devoted to the illustration of
the general concepts in the context of the large-$N$ model and of the $XY$ model, respectively. Conclusions are
presented in section~\ref{sec8}.

\section{Scalings of $C(t,t_w)$ and $R(t,t_w)$} \label{sec2}

In the following we consider quenches from an initial disordered state above the critical line 
to any one of the states in the shaded area of Fig.1. For definiteness, we shall refer to 
purely relaxational dynamics without 
conservation of the order parameter, but the results are of general validity.  The common feature of all these
processes is that, after a short transient, dynamical scaling sets in with a characteristic length
growing with the power law $L(t) \sim t^{1/z}$~\cite{Bray}.
The scaling properties with $T=T_C$ are different from those with $T<T_C$~\cite{Godreche}.
In the renormalization group language this means that for a given $d$ 
there are two fixed points, one unstable at $T_C$ and the other stable and
attractive at $T=0$ (thermal fluctuations are irrelevant below $T_C$)~\cite{Mazenko85}.
The dynamical exponent $z$ on the $T_C(d)$ line 
coincides with the dynamical critical exponent $z_c$ and depends on dimensionality, while below criticality 
it takes the dimensionality independent value $z=2$. These values become identical at $(d_L,T=0)$, since
$\lim_{d \to d_L} z_c = 2$~\cite{Hohenberg}.

Let us, next,  review the scaling properties of  $C(t,t_w)$ and $R(t,t_w)$. 
The fixed point structure enters in two ways: in the values of the exponents involved and, less obviously,
in the very same form of the functions $C(t,t_w)$ and $R(t,t_w)$. 
In analysing the asymptotic behavior of these quantities, it is necessary to distinguish between the
{\it short time} and the {\it large time} behaviors obtained by letting the waiting time
$t_w$ to become large, while  keeping, respectively, either the time difference $\tau=t-t_w$ or 
the time ratio $x=t/t_w >1$ fixed. Notice that from
$x = 1- \tau/t_w$ follows that the short time regime, when using the $x$ variable, gets compressed into $x=1$.

An important general property of two time quantities in slow relaxation phenomena
is that in the short time regime 
the system appears equilibrated and the FDT is satisfied, while the off equilibrium character of the dynamics, or
aging, shows up in the large time regime~\cite{Cugliandolo2002,Crisanti,Calabrese}. 
The mechanism underlying this property, however, is quite
different for quenches to and below $T_C$.

\subsection{Critical quench: multiplicative structure}  \label{sec2.1}

In the critical quenches, for $t_w$ sufficiently large, $C(t,t_w,t_0)$ and  $R(t,t_w,t_0)$ take
the product forms~\cite{Godreche,Calabrese}
\be
C(t,t_w,t_0) = (\tau +t_0)^{-b}g_C(x,y)
\label{SCR1}
\ee
\be
R(t,t_w,t_0) = (\tau +t_0)^{-(1+a)}g_R(x,y)
\label{SCR2}
\ee
where $g_{C,R}(x,y)$ are smooth functions of $x$ with a weak dependence on $y=t_0/t_w$.
In addition to the 
observation times, $t$ and $t_w$, we have explicitely included also the 
dependence on a microscopic time $t_0$, which is needed to regularize these functions at equal times.
In particular, we shall take $C(t,t,t_0)=1$ throughout.
In the short time regime
\be
C(t,t_w,t_0) = (\tau +t_0)^{-b}g_C(1,0)
\label{SCR5}
\ee
\be
R(t,t_w,t_0) = (\tau +t_0)^{-(1+a)}g_R(1,0)
\label{SCR6}
\ee
which are the autocorrelation and response function of the stationary critical dynamics
satisfying, therefore, the equilibrium FDT, i.e. $T_C R(\tau) = -\partial_{\tau} C(\tau)$, 
which implies
\be
T_C g_R(1,0)= b g_C(1,0)
\label{S6}
\ee
and
\be
a=b.
\label{S60}
\ee
We emphasize that, in the critical quench, as a consequence of the multiplicative structure and of the 
FDT, the exponents $a$ and $b$ are not independent and coincide. Furthermore, their common value is
given by~\cite{Godreche,Calabrese}
\be
a=b=(d-2+\eta)/z_c = 2\beta/\nu z_c
\label{207}
\ee
where $\eta$, $\beta$ and $\nu$ are the usual static exponents. Using the geometrical interpretation
of the critical properties Eq.~(\ref{207}) can be rewritten as $a=b=2(d-D)/z_c$, where $D$ is the fractal
dimensionality of the correlated critical clusters~\cite{Coniglio}. These become compact as $T_C \rightarrow 0$
yielding
\be
\lim_{d \to d_L} a=b=0.
\label{207.1}
\ee
On the KT line, where $d=2$, Eq.~(\ref{207}) is replaced by
$a=b=\eta(T)/z$ with $\eta(T)$ vanishing as $T \rightarrow 0$~\cite{kosterlitz}.
Eqs.~(\ref{SCR1}) and~(\ref{SCR2}) can be recast in the simple aging form
\be
C(t,t_w,t_0) = t_w^{-b}f_C(x,y) 
\label{SCR7}
\ee
\be
R(t,t_w,t_0) = t_w^{-(1+a)}f_R(x,y) 
\label{SCR8}
\ee
where
\be
f_C(x,y)= (x-1+y)^{-b}g_C(x,y)
\label{SCR9}
\ee
\be
f_R(x,y)= (x-1+y)^{-(1+a)}g_R(x,y).
\label{SCR9bis}
\ee
These functions, for large  $x$ and $t_w$, decay with the same power law
\be
f_{C,R}(x,0) = A_{C,R} x^{-\lambda_c/z_c}
\label{SCR10}
\ee
where $\lambda_c$ is the autocorrelation exponent~\cite{Godreche,Calabrese}.

\subsection{Quench below the critical line: additive structure} \label{sec2.2}

Below $T_C$, the picture is different because the stationary and aging components 
enter {\it additively} into the autocorrelation function~\cite{Bouchaud97,Cugliandolo2002,Crisanti}
\be
C(t,t_w,t_0) = C_{st}(\tau)+  C_{ag}(t,t_w,t_0) 
\label{4.01}
\ee
where $C_{st}(\tau)$ is the equilibrium autocorrelation function in the broken symmetry pure state
at the temperature $T$. 
From the equal time properties
\be
C(t,t,t_0)=1 \qquad C_{st}(\tau=0)=1-M^2  
\label{4.01bis}
\ee
which imply
\be
C_{ag}(t,t,t_0)=M^2
\label{4.01quater}
\ee
where $M$ is the spontaneous magnetization,
follows that in Eq.~(\ref{4.01}) the stationary component corresponds to the quasi-equilibrium decay {\it to} the plateau at 
the Edwards-Anderson order parameter $q_{EA}=M^2$, while the aging component $C_{ag}$ describes
the off-equilibrium decay {\it from} the plateau at much larger times.
In the short time regime, then, we have
\be
C(\tau +t_w,t_w,t_0) = C_{st}(\tau) + M^2
\label{SCR15}
\ee
while, taking into account that $C_{st}(\tau)$ vanishes as $\tau \rightarrow \infty$,
in the large time regime 
\be
C(t,t_w,t_0) = C_{ag}(t,t_w,t_0).
\label{SCR16}
\ee
From this it is easy to verify that, contrary to what happens with the multiplicative structure of
Eq.~(\ref{SCR1}), the
limits $t_w \rightarrow \infty$ and $t \rightarrow \infty$ do not commute
\be
\lim_{t \to \infty} \lim_{t_w \to \infty} C(t,t_w,t_0) = M^2
\label{SCR17}
\ee
\be
\lim_{t_w \to \infty}\lim_{t \to \infty}C(t,t_w,t_0) = 0
\label{SCR18}
\ee
yielding the weak ergodicity breaking~\cite{Cugliandolo2002}, characteristic of the quenches below $T_C$.

The additive structure of the response function 
\be
R(t,t_w,t_0) = R_{st}(\tau)+  R_{ag}(t,t_w,t_0)
\label{4.02}
\ee
is obtained in the following way~\cite{Crisanti}: in the short time regime stationarity holds $R(t,t_w,t_0)=R_{st}(\tau)$
and $R_{st}(\tau)$ is defined from the FDT 
\be
R_{st}(\tau) =- {1 \over T} {\partial \over \partial \tau}C_{st}(\tau)
\label{STFDT}
\ee  
whilst
$R_{ag}(t,t_w,t_0)$ remains defined by Eq.~(\ref{4.02}). This implies that $R_{ag}(t,t_w,t_0)$
vanishes in the short time regime and since, conversely, $R_{st}(\tau)$ vanishes in the large time regime,
we may write
\begin{eqnarray} 
R(t,t_w,t_0) =   \left \{ \begin{array}{ll}
        R_{st}(\tau)  \qquad $for$ \qquad x = 1 \\
        R_{ag}(t,t_w,t_0)   \qquad $for$ \qquad  x > 1.
        \end{array}
        \right .
        \label{R1}
        \end{eqnarray} 
The components $C_{ag}(t,t_w,t_0)$ and $R_{ag}(t,t_w,t_0)$ obey simple aging forms like~(\ref{SCR7}) 
and~(\ref{SCR8}) with scaling functions
$f_{C,R}(x,y)$ decaying, for large $x$, with a power law of the form~(\ref{SCR10}). It is understood that all the exponents 
must be replaced by their values below $T_C$, which are different from those at $T_C$. 
In particular, from phase ordering theory it is well known~\cite{Bray} that $b=0$ independently
from dimensionality. This can be readily understood from $b=2(d-D)/z$, which holds in general, 
and from the compact nature of the coarsening domains, which gives $D=d$. 
Quite different is the case of the exponent $a$. Since the FDT relates only the stationary
components $C_{st}(\tau)$ and $R_{st}(\tau)$, due to the additive structure of Eqs.~(\ref{4.01}) and~(\ref{4.02})
we have no more the constraint $a=b$, as in the critical quench. Namely, in the quench below $T_C$
the value of $a$ is decoupled from that of $b$. As a matter of fact, the determination of $a$ below $T_C$ is a difficult
and challanging problem~\cite{Generic,Corberi2003}, which will be discussed in section~\ref{sec5}.

\subsection{Quench at $(d_L,T=0)$: additive structure} \label{sec2.3}

As we have seen above, the structures of $C(t,t_w,t_0)$ and $R(t,t_w,t_0)$ follow different patterns
in the quenches to and below $T_C$. This raises the question of 
which one applies when the quench is made at $(d_L,T=0)$. In other words, should we consider 
the state $(d_L,T=0)$ as a critical
point with $T_C=0$, as it is often done in the literature, or should we consider it as the continuation to $d_L$ of the
$T=0$ states below criticality? The correct answer is the second one, since at $d_L$ the system orders,
as in all the states below the critical line, while on the $T_C(d)$ line there is no ordering.
In particular, this implies that weak ergodicity breaking takes place also in the quench to $(d_L,T=0)$.
From Eqs.~(\ref{4.01}) and~(\ref{4.01bis}) then follows $C_{st}(\tau) \equiv 0$ and 
\be
C(t,t_w,t_0) = f_C(x,y)
\label{SCR19}
\ee
with $f_C(1,0)=1$, since $M^2=1$ at $T=0$. 

The fact that the quench to $(d_L,T=0)$ belongs to the additive scheme becomes important
when considering the response function, because if $C_{st}(\tau)$ vanishes in the ground state, the 
same is not true for $R_{st}(\tau)$. Therefore, Eq.~(\ref{4.02}) holds with $R_{st}(\tau) \neq 0$
and it is important to subtract this contribution from $R(t,t_w,t_0)$ if one wants to study
the scaling properties of $R_{ag}(t,t_w,t_0)$. We shall come back on this point in section~\ref{sec6},
when treating the explicit example of the large-$N$ model.
It should be mentioned, here, that the linear response function at $T=0$ is well defined only
for systems with soft spins. In the case of hard spins there does not exist a linear response regime when 
$T=0$. For a discussion of this problem and how to bypass it in Ising systems, we refer to~\cite{Lippiello2000,Corberi2001}.

\subsection{Comparison with theory}

A natural question is how the picture outlined above 
compares with theoretical approaches. In the case of the quench to $T_C$ renormalization group 
methods are available~\cite{Janssen,Calabrese}, which account for the multiplicative structure of
the autocorrelation and response function and give explicit expressions for the quantities of
interst up to two loops order. The development of systematic expansion methods for the quench below $T_C$ is
much more difficult~\cite{Mazenko}.
Recently, Henkel and collaborators~\cite{Henkel,henkeletal} have developed a method based on the
requirement of local scale invariance. According to this approach 
the response function obeys the multiplicative structure~(\ref{SCR2}) {\it both} in the quenches to
$T_C$ and below $T_C$. Problems, then, arise in the latter case. The main one is of conceptual
nature, since, as explained above, the weak ergodicity breaking scenario requires the autocorrelation
function to obey the additive structure~(\ref{4.01}). Then, in the short time regime, where 
Eq.~(\ref{SCR15}) holds, from the FDT one has $\partial_{t_w} C_{st}(\tau) \sim (\tau + t_0)^{-(1+a)}$, which implies
$ C_{st}(\tau) \sim (\tau + t_0)^{-a}$. In other words, if the local scale invariance prediction for
the response function is valid then the stationary correlation
function must decay with a power law. This is the case, as we have seen above, in the critical 
quenches and, as we shall see
in section~\ref{sec6}, in the large $N$ model. Conversely, in all other cases
where the equilibrium correlation function below $T_C$ decays exponentially, the response function cannot be
of the form~(\ref{SCR2}). Nonetheless, Henkel et al. do
criticize~\cite{henkeletal} the additive structure below $T_C$.

The second problem is that existing results, exact~\cite{Lippiello2000,Godreche2000,Corberi2002} 
and approximated~\cite{Corberi2001,Berthier99,Mazenko} as well as numerical~\cite{Corberi2003,algorithm}, lead to 
an the aging part of the response function which, for systems with scalar order parameter~\cite{vector}, is of the form
\be
R_{ag}(t,t_w,t_0) = t_w^{-1/z}(\tau + t_0)^{-(a-1/z+1)}g_R(x,y).
\label{rag}
\ee
This is qualitatively different from the multiplicative form~(\ref{SCR2}), both in the structure and in the exponents.
As mentioned above, an exception is the large $N$ model, where $R(t,t_w,t_0)$ is of the
form~(\ref{SCR2}) also below $T_C$. Yet, as we show in section~\ref{sec6}, the additive structure applies to the large $N$
model as in all cases of quench below $T_C$.

\section{Effective temperature} \label{sec3}

The next step is to see how these different behaviors of $C(t,t_w,t_0)$ and $R(t,t_w,t_0)$ do
affect the behavior of $T_{eff}(t,t_w,t_0)$.

\subsection{Critical quench} \label{sec3.1}

From Eqs.~(\ref{SCR7}),~(\ref{SCR8}) and~(\ref{S60}), 
using the definition~(\ref{0.1bis}), we get
\be
T_{eff}(t,t_w,t_0)  = F(x,y)
\label{202}
\ee
with
\be
F(x,y) =  - f_{\partial C}(x,y)/f_R (x,y) 
\label{203}
\ee
and
\be
f_{\partial C}(x,y) = b f_C(x,y) +\left [ x { \partial \over \partial x}f_C(x,y)
+ y { \partial \over \partial y}f_C(x,y) \right ].
\label{204}
\ee
In the short time regime, Eqs.~(\ref{SCR5}),~(\ref{SCR6}) and~(\ref{S6}) lead to the boundary condition
\be
F(1,y)=T_C
\label{EF1}
\ee
while, in the aging regime, from Eq.~(\ref{SCR10}) follows 
\be
\lim_{x \to \infty} F(x,0)= T_{\infty} = \left ({\lambda_c \over z_c} -b \right ) { A_C \over A_R}
= {T_C \over X_{\infty}}
\label{EFT1}
\ee
where $X_{\infty}$ is the limit FDR of Godr\`eche and Luck~\cite{Godreche}.

\subsection{Quench below the critical line} \label{sec3.2}

Below $T_C$, where $b=0$, from the definition~(\ref{0.1bis}), the additive structure and Eq.~(\ref{STFDT}) we get
\be
T_{eff}(t,t_w,t_0)) = {T R_{st}(\tau) \over R_{st}(\tau) + R_{ag}(t,t_w,t_0)} +
{\partial_{t_w}C_{ag}(t,t_w,t_0) \over R_{st}(\tau) + R_{ag}(t,t_w,t_0)}
\label{EF2}
\ee
which can be rewritten as
\begin{eqnarray}
T_{eff}(t,t_w,t_0)) & = &  T \left [ { R_{st}(\tau) \over R_{st}(\tau) + R_{ag}(t,t_w,t_0)} \right ] \label{0003}
  \\
& + & t_w^a F(x,y)  \left [ { R_{ag}(t,t_w,t_0) \over R_{st}(\tau) + R_{ag}(t,t_w,t_0)} \right ] \label{pippooo} \nonumber 
\end{eqnarray}
where $F(x,y)$ is still defined by  Eqs.~(\ref{203}),~(\ref{204}) with $f_C(x,y)$, $f_R(x,y)$  
the scaling functions of $C_{ag}(t,t_w,t_0)$ and $R_{ag}(t,t_w,t_0)$. 

Using Eq.~(\ref{R1}) we may approximate Eq.~(\ref{0003}) with
\be
T_{eff}(t,t_w,t_0)) = T H(x) + t_w^a F(x,y) [1-H(x)]
\label{PQR}
\ee
where
\begin{eqnarray}
H(x) =   \left \{ \begin{array}{ll}
        1   \qquad $for$ \qquad x = 1 \\
        0   \qquad $for$ \qquad  x > 1.
        \end{array}
        \right .
        \label{KK}
        \end{eqnarray}
Therefore, for large $t_w$ the general formula containing both critical and subcritical quenches is given by~\cite{note}
\begin{eqnarray}
T_{eff}(t,t_w,t_0)) =   \left \{ \begin{array}{ll}
        T  \qquad $for$ \qquad x = 1 \\
        t_w^{a-b}F(x,y)  \qquad $for$ \qquad  x > 1.
        \end{array}
        \right .
        \label{PR1}
        \end{eqnarray}
Of course, this includes also the quench to $(d_L,T=0)$.

\section{Parametric representation} \label{sec4}

In order to derive the parametric representation, as explained in the Introduction, we must express the time
dependence of $T_{eff}(t,t_w,t_0)$ through  $C(t,t_w,t_0)$ and then let $t_w \rightarrow \infty$, 
obtaining ${\cal T}(C)$.
From Eqs.~(\ref{SCR1}) and~(\ref{4.01}), for large $t_w$, the general form of the autocorrelation function can be written as
\begin{eqnarray}
C(t,t_w,t_0) =   \left \{ \begin{array}{ll}
        1  \qquad $for$ \qquad x = 1 \\
        t_w^{-b}f_C(x,y)   \qquad $for$ \qquad  x > 1.
        \end{array}
        \right .
        \label{PR2}
        \end{eqnarray}
The task is to eliminate
$x$ between Eqs.~(\ref{PR1}) and~(\ref{PR2}), which now we do separately for the quenches into the
different regions of the phase diagram.

\subsection{Critical quench} \label{sec4.1}

In this case  $a=b>0$.
From Eqs.~(\ref{SCR7}) and~(\ref{SCR9}), letting $t_w \rightarrow \infty$ with $x$ fixed, we obtain the singular limit
\begin{eqnarray}
C(x) =  \left \{ \begin{array}{ll}
        1  \qquad $for$ \qquad x=1 \\
        0   \qquad $for$ \qquad x > 1
        \end{array}
        \right .
        \label{G1}
        \end{eqnarray}
whose inverse is readily obtained exchanging the horizontal with the vertical axis
\begin{eqnarray}
x(C) =  \left \{ \begin{array}{ll}
        \infty \qquad $for$ \qquad C=0 \\
        1   \qquad $for$ \qquad 0 < C \leq 1 .
        \end{array}
        \right .
        \label{EFT4}
        \end{eqnarray}
Inserting into Eq.~(\ref{PR1}) we obtain (Fig.2a)
\begin{eqnarray} 
{\cal T}(C) =   \left \{ \begin{array}{ll}
        T_{\infty}> T_C  \qquad $for$ \qquad C =0\\
        T_C   \qquad $for$ \qquad   0 <C \leq 1
        \end{array}
        \right .
        \label{I1}
        \end{eqnarray} 
where we have used Eqs.~(\ref{EF1}) and~(\ref{EFT1}). 
This is a universal result, since all the non universal features of $F(x,0)$ for the intermediate values of $x$ 
have been eliminated in the limit process.
Therefore, we have that for all quenches to $T_C >0$,
apart for the value $T_{\infty}$ at $C=0$, the parametric plot of ${\cal T}(C)$ is trivial in the sense that
the effective temperature coincides with the temperature $T_C$ of the thermal bath.   
Notice that this implies that for the ZFC susceptibility
\be
\chi(t,t_w) = \int_{t_w}^t ds R(t,s)
\label{chi}
\ee
whose parametric form is related to ${\cal T}(C)$ by
\be
\chi(C)= \int_C^1 {dC' \over {\cal T}(C')}
\label{XY3}
\ee
a linear plot is obtained
\be
\chi(C)= (1-C)/T_C
\label{XY3bis}
\ee
as for equilibrated systems~\cite{Crisanti}.

\subsection{Quench below the critical line}  \label{sec4.2}

In this case $b=0$ and from Eq.~(\ref{PR2}) follows
\begin{eqnarray}
C(x) =   \left \{ \begin{array}{ll}
        1  \qquad $for$ \qquad x = 1 \\
        f_C(x,0)   \qquad $for$ \qquad  x > 1
        \end{array}
        \right .
        \label{PR3}
        \end{eqnarray}
where, recalling Eqs.~(\ref{4.01bis}) and~(\ref{SCR16}), $f_C(x,0)$ is the smooth function describing the fall below the plateau
at $M^2$. Therefore, the inverse function is given by
\begin{eqnarray}
x(C) =  \left \{ \begin{array}{ll}
        f_C^{-1}(C)>1 \qquad $for$ \qquad C < M^2 \\
        1   \qquad $for$ \qquad M^2 \leq C \leq 1
        \end{array}
        \right .
        \label{PR4}
        \end{eqnarray}
and inserting into Eq.~(\ref{PR1}) we find
\begin{eqnarray}
\widehat{T}_{eff}(C,t_w) =   \left \{ \begin{array}{ll}
        t_w^{a}F( f_C^{-1}(C),0)   \qquad $for$ \qquad  C < M^2 \\
        T   \qquad $for$ \qquad M^2 \leq C \leq 1.
        \end{array}
        \right .
        \label{G6bis}
        \end{eqnarray}
Although the actual value of $a$, as will be explained in section~\ref{sec5}, is to some extent a debated issue, 
there is consensus on $a>0$
for $d>d_L$. Therefore, taking the  $t_w \rightarrow \infty$ limit we recover the well known 
result~\cite{Cugliandolo2002,Crisanti} (Fig.2b)
\begin{eqnarray} 
{\cal T}(C) =   \left \{ \begin{array}{ll}
        \infty \qquad $for$ \qquad C < M^2\\
        T   \qquad $for$ \qquad M^2  \leq C \leq 1.
        \end{array}
        \right .
        \label{I2}
        \end{eqnarray} 
Again, all the details of $F(x,0)$ having disappeared, the function
${\cal T}(C)$ is universal.

Let us remark, here, as a further manifestation of the difference between quenches to and below $T_C$,
that the behavior~(\ref{I1}) of ${\cal T}(C)$ cannot be recovered by letting
$T \rightarrow T_C^-$ and $M \rightarrow 0$ in~(\ref{I2}). In fact, in the latter case we find
\begin{eqnarray} 
\lim_{T \to T_C^-} {\cal T}(C) =   \left \{ \begin{array}{ll}
        \infty \qquad $for$ \qquad C =0\\
        T_C   \qquad $for$ \qquad 0 < C \leq 1
        \end{array}
        \right .
        \label{I2bis}
        \end{eqnarray}
which differs from Eq.~(\ref{I1}), where $T_{\infty}$ is a finite number.

\subsection{Quench to $(d_L,T=0)$} \label{sec4.3}

If we take the limit $d \rightarrow d_L$ in Eqs.~(\ref{I1}) and~(\ref{I2}) we obtain, respectively
\begin{eqnarray} 
{\cal T}(C) =   \left \{ \begin{array}{ll}
        T_{\infty}  \qquad $for$ \qquad C =0\\
        0   \qquad $for$ \qquad   0 <C \leq 1
        \end{array}
        \right .
        \label{I1.10}
        \end{eqnarray} 
and
\begin{eqnarray} 
{\cal T}(C) =   \left \{ \begin{array}{ll}
        \infty \qquad $for$ \qquad C < 1\\
        0   \qquad $for$ \qquad  C = 1
        \end{array}
        \right .
        \label{I2.11}
        \end{eqnarray}
which are very much different one from the other. 
So, there is the problem of which is the form of ${\cal T}(C)$ in the quench to $(d_L,T=0)$.
A statement to this regard can be made on the basis of the substantial amount of information by now 
accumulated from sources as diverse as the exact solutions of the $1d$ Ising model~\cite{Lippiello2000,Godreche2000} and of the
$2d$ large-$N$ model~\cite{Corberi2002}, the Ohta-Jasnow-Kawasaki type of
approximations with $d=1$~\cite{Corberi2001}, in addition to numerical simulations~\cite{Generic} of systems
at $d_L$ with scalar and vector order parameter, with and without conserved dynamics. In
all of these cases we have obtained the parametric plot~(\ref{XY3}) of the ZFC susceptibility,
which gives for ${\cal T}(C)$ at $(d_L,T=0)$ a non trivial and non universal finite function 
of the type depicted in Fig.2c. By contrast, there does not exists, up to now, any evidence for behaviors of ${\cal T}(C)$
of the type~(\ref{I1.10}) or~(\ref{I2.11}).

Assuming that the generic behavior of ${\cal T}(C)$ is of the type in Fig.2c, as the amount of evidence quoted above
strongly suggests,
this is compatible with Eq.~(\ref{PR1}) only if $a=0$. Then, from Eqs.~(\ref{PR3}),~(\ref{PR4}) and~(\ref{G6bis})
with $M^2=1$ we may write
\begin{equation} 
{\cal T}(C) = F(f_C^{-1}(C),0)
        \label{EF8}
        \end{equation}
which represents a smooth, finite, non trivial function deacreasing smoothly from $T_{\infty}= F(\infty,0)$ at
$C=0$ toward zero at $C=1$, preserving all the non universal features of 
$f_C(x,0)$ and $F(x,0)$ for the intermediate values of $C$.

\subsection{Physical temperature and connection with statics} \label{sec4.4}

We conclude this section with comments on the interpretation of ${\cal T}(C)$ as a true temperature and on
its connection with the equilibrium properties of the system. 

The relation of the effective temperature with a physically measurable temperature is a very interesting
issue, which has been thoroughly investigated in Refs.~\cite{Peliti,Sollich,CG}. A necessary condition for the identification of
the effective temperature with a true temperature is its uniqueness, that is the indepence from the
observable used in the study of the deviation from FDT. For what concerns the quenches to $T_C$, in~\cite{CG}
it is shown that the multiplicative structure~(\ref{SCR1}) and~(\ref{SCR2}) is observable independent.
Therefore, the derivation of Eq.~(\ref{I1}) is also observable independent, except for the actual 
value of $T_{\infty}$, which does to depend on the observable~\cite{CG}. In this case, the criterion
of uniqueness does indeed lead to the identification of the effective temperature with a physical
temperature, since ${\cal T}(C)$, for $C>0$, coincides with the temperature of the thermal bath $T_C$.
Conversely $T_{\infty}$, due to the non uniqueness of its value, cannot be considered as physical temperature.
Although results of comparable generality are lacking for the quenches below $T_C$, we expect
also Eq.~(\ref{I2}) to be observable independent, due to the universality of ${\cal T}(C)$ (for observables with $a>0$). 
Finally, the effective temperature~(\ref{EF8}), in the quench to $(d_L,T=0)$, it is certainly not 
physical, since, as explained above, it does to depend on the details of the scaling function, which
makes it observable dependent. To this case belong the results for the $d=1$ Ising model~\cite{Sollich}, which do indeed
show the observable dependence of the effective temperature.   

The connection between static and dynamic properties is given by Eq.~(\ref{fmpp}).
We can now assess the status of this relation, in the light of the results derived above.
In all cases, the overlap probability function is given by $P(q)= \delta (q-M^2)$.
We must, then, verify if the r.h.s. of Eq.~(\ref{fmpp}) actually is a $\delta$
function. From Eqs.~(\ref{I1},\ref{I2},\ref{EF8}) follows that $T {d \over dC} {\cal T}(C)^{-1}|_{C=q}$
vanishes everywhere, except at $C=M^2$. Therefore, in order to establish whether this is a $\delta$
function, we must look at the integral $\int_0^1 dC {d \over dC}{T \over {\cal T}(C)}$.
This gives finite values  in the quenches to $T_C$
and below $T_C$. The validity of Eq.~(\ref{fmpp}), instead, cannot be established in the quench to $(d_L,T=0)$.
In that case the value of the integral is not determined, since the contribution at the upper limit of integration is given by the ratio of two vanishing quantities.

\section{The response function exponent} \label{sec5}

With the results for ${\cal T}(C)$ illustrated in the previous section we come to grips with the problem
of the exponent $a$. The vanishing of $a$, required by the behavior of ${\cal T}(C)$ in the quench to
$(d_L,T=0)$, cannot be accounted for
by Eq.~(\ref{207.1}) because, as explained in section~\ref{sec2.3}, the quench to $(d_L,T=0)$ is not
a critical quench. Hence, the vanishing of $a$ must be framed within the behavior of $a$ in the quenches below
the critical line.
 
According to a popular conjecture~\cite{Barrat}, $a$ for $T < T_C$ should coincide with the exponent $n/z$ entering in the
time dependence of the density of defects, which goes like $t^{-n/z}$ with $n=1$ or $n=2$ for scalar or vector order parameter,
respectively~\cite{Bray}. We recall that for $T<T_C$ the dynamical exponent $z=2$ does not depend on $d$. 
Thus, in the scalar case one ought to have $a=1/2$ and in the vector case $a=1$,
{\it independently} from $d$ and, therefore, also at $d_L$. This is obviously incompatible with 
the large body of evidence for $a=0$ at $d_L$, quoted in the previous section.

This difficulty is supersed in the alternative proposal
which we have made~\cite{Corberi2001,Corberi2003} for the behavior of $a$ as $d$ is varied, on the basis of 
the exact solution of the large-$N$ model~\cite{Corberi2002}
for arbitrary $d$ and the Ohta-Jasnow-Kawasaki type of approximation~\cite{Corberi2001,Berthier99,Mazenko}, 
also for arbitrary $d$. In these two cases
one finds that $a$ does depend on $d$ according to
\be
a={n \over z} \left ({d-d_L \over d_U - d_L} \right )
\label{4.2}
\ee 
where $d_U$ is a parameter dependent on the universality class.
We have proposed to promote this result to a phenomenological formula, whose general validity we have
tested by undertaking a systematic numerical investigation of systems in different classes of universality and at different 
dimensionalities~\cite{Corberi2001,Generic,Corberi2003}. The results obtained are in quite good agreement with Eq.~(\ref{4.2}), taking
$d_U =3$ or $d_U =4$, respectively for scalar~\cite{note1} or vector order parameter. We emphasize that
Eq.~(\ref{4.2}) gives $\lim_{d \to d_L} a=0$ when the limit is taken {\it below} $T_C$, and this should not be confused with
Eq.~(\ref{207.1}) holding for the critical quenches.

The final remark is that, up to now  Eq.~(\ref{4.2}) remains a phenomenological formula. The challange is to make
a theory for it. A preliminary attempt to relate the dimensionality dependence of $a$ to the roughening of interfaces in the
scalar case has been made in~\cite{Generic}. However, the explanation of Eq.~(\ref{4.2}) as a result of general
validity seems to require an understanding of the response function much deeper than we presently have.

\section{Large-$N$ model} \label{sec6}

In this section all the concepts introduced above are explicitely illustrated through the exact solution of 
the large-$N$ model. This model, or the equivalent spherical model, has been solved analytically in a 
number of papers~\cite{Cugliandolo95,Godreche-sph,Zippold}.
Here, we follow our own solution in Ref.~\cite{Corberi2002},  where we have shown that for quenches to $T
\leq T_C$ the order parameter can be split into the sum of two stochastic processes
$\phi(\vec{x},t) = \sigma(\vec{x},t) + \psi(\vec{x},t)$ which, for $t$ sufficiently large, are independent 
and mimic the separation between the interface and bulk fluctuations
taking place in domain forming systems~\cite{Franz2000}.
Correspondingly, the autocorrelation function decomposes into the sum of the autocorrelations of $\sigma$ and $\psi$ 
\be
C(t,t_w,t_0) = C_{\sigma}(t,t_w,t_0) + C_{\psi}(t,t_w,t_0)
\label{N1}
\ee
which obey the scaling forms
\be
C_{\sigma}(t,t_w,t_0) = t_w^{-b_{\sigma}} f_{\sigma}(x,y) 
\label{N1bis}
\ee
and
\be
C_{\psi}(t,t_w,t_0) =  t_w ^{-b_{\psi}}f_{\psi}(x,y).
\label{N1tris}
\ee
For the purposes of the present paper it is sufficient to limit the analysis to $d < 4$, where the exponents and the 
scaling functions are given by
\begin{eqnarray}
b_{\sigma} =  \left \{ \begin{array}{ll}
        d - 2 \qquad $for $ \qquad  T=T_C  \\
        0  \qquad $for$ \qquad T < T_C
        \end{array}
        \right .
        \label{N6}
        \end{eqnarray}
\be
b_{\psi} =(d-2)/2
\label{N7}
\ee
\be
f_{\sigma}(x,y) = A_{\sigma}x^{-\omega/2} \left [ {1 \over 2} (x+1+y) \right ]^{-d/2}
\label{N4}
\ee
\be
f_{\psi}(x,y) = {2T \over (4 \pi)^{d/2}}x^{-\omega/2}
\int_0^1 dz z^{\omega}  (x+1-2z+y)^{-d/2}
\label{N5}
\ee
\begin{eqnarray}
\omega =  \left \{ \begin{array}{ll}
        d/2 - 2 \qquad $for $ \qquad  T=T_C  \\
        -d/2  \qquad $for$ \qquad T < T_C
        \end{array}
        \right .
        \label{N8}
        \end{eqnarray}
and $A_{\sigma}$ is a constant, which coincides with $M^2$ in the quenches below $T_C$.
The scaling function $f_{\psi}(x,y)$ can also be rewritten as
\be
f_{\psi}(x,y)=(x-1+y)^{-b_{\psi}}g_{\psi}(x,y)
\label{N5bis}
\ee
with
\be
g_{\psi}(x,y)= {T \over (4 \pi)^{d/2}}x^{-\omega/2}
\int_0^{2/( x-1+y)} dz (1+z)^{-d/2} \left [1- {1 \over 2} (x-1+y)z \right ]^{\omega}
\label{N5tris}
\ee
which allows to recast $C_{\psi}(t,t_w,t_0)$ in the multiplicative form
\be
C_{\psi}(t,t_w,t_0)= (\tau +t_0)^{-b_{\psi}}g_{\psi}(x,y).
\label{N51tris}
\ee

The response function is given by 
\be
R(t,t_w,t_0) = t_w^{-(1+a)}f_R(x,y)
\label{N9}
\ee
with
\be
f_R(x,y) = (4 \pi)^{-d/2}(x-1+y)^{-(1+a)}x^{-\omega/2}
\label{N9tris}
\ee
and
\be
a= b_{\psi} = (d-2)/2.
\label{N9bis}
\ee
Finally, in the phase diagram of Fig.1 the critical line is a straight line 
\be
T_C= (4\pi)^{d/2} \Lambda^{2-d} (d-d_L)/2
\label{ScN1}
\ee
where  $\Lambda$ is the momentum cutoff and $d_L=2$.

\subsection{Scalings of $C(t,t_w,t_0)$ and $R(t,t_w,t_0)$} \label{sec6.1}

Let us now extract from the above results the scaling  properties of $C(t,t_w,t_0)$ and $R(t,t_w,t_0)$ 
on the different regions of the phase diagram.

\begin{enumerate}

\item {\it Critical quench}  

The first observation is that
the autocorrelation and response function have the multiplicative
structures~(\ref{SCR1}) and~(\ref{SCR2}). For $R(t,t_w,t_0)$ this is immediately evident from Eq.~(\ref{N9tris}),
where $f_R(x,y)$ is of the form~(\ref{SCR9bis}) with  $g_R(x,y)= (4 \pi)^{-d/2}x^{1-d/4}$.

For $C(t,t_w,t_0)$, from $b_{\sigma}= 2b_{\psi} $ follows that for large $t_w$
the first term in Eq.~(\ref{N1}) is negligible with respect to the second one, yielding
\be
C(t,t_w,t_0) = C_{\psi}(t,t_w,t_0)
\label{NN1}
\ee
and this is of the form~(\ref{SCR1}) and~(\ref{SCR7}) with
the identifications $b=b_{\psi}$, $f_C(x,y)=f_{\psi}(x,y)$
and $g_C(x,y)=g_{\psi}(x,y)$.
Furthermore, from Eqs.~(\ref{N7}) and~(\ref{N9bis}) follows
\be
a=b=(d-2)/2
\label{NN3}
\ee
in agreement with Eq.~(\ref{207}), since in the large-$N$ model $\eta=0$ and $z_c=2$. Finally,
from Eq.~(\ref{N5tris}) we have $g_{C}(1,0)=(4\pi)^{-d/2}2T_C/(d-2)$ and, using 
$g_{R}(1,0)= (4\pi)^{-d/2}$, it is
easy to check that Eq.~(\ref{S6}) is verified.

\item {\it Quench below the critical line}

For large $t_w$, the additive structure of Eq.~(\ref{4.01}) is found ready made in Eq.~(\ref{N1}),
with the identifications
\be
C_{st}(\tau) = C_{\psi}(\tau)= (\tau +t_0)^{-b_{\psi}}g_{\psi}(1,0)
\label{NN4}
\ee
where $g_{\psi}(1,0)=(4\pi)^{-d/2}2T/(d-2)$ and 
\be
C_{ag}(t,t_w,t_0) = C_{\sigma}(t,t_w,t_0)
\label{NN6}
\ee
which gives $b=b_{\sigma}=0$ and $f_C(x,y)=f_{\sigma}(x,y)$. The pawer law decay~(\ref{NN4}) of the stationary component
$C_{st}(\tau)$ is a peculiarity of the large-$N$ model, since below $T_C$ to lowest order in $1/N$ only the
Goldstone modes contribute to thermal fluctuations~\cite{Mazko-Z}.

Carrying out the prescription outlined in section~\ref{sec2.2} for the construction of the corresponding components of the response function,
from Eq.~(\ref{STFDT}) we have
\be
R_{st}(\tau) = (4 \pi)^{-d/2}(\tau+t_0)^{-(1+b_{\psi})}
\label{NN7}
\ee
and using the identity $b_{\psi}=a$ we find
\begin{eqnarray}
R_{ag}(t,t_w,t_0) & = & R(t,t_w,t_0) - R_{st}(\tau) \\ \nonumber
& = & t_w^{-(1+a)}f_R(x,y)      
\label{NN8}
\end{eqnarray}
with
\be
f_R(x,y)=(4 \pi)^{-d/2}{x^{d/4}-1 \over (x-1+y)^{1+a}}.
\label{NN8bis}
\ee

\item {\it Quench to} $T =0$ {\it with} $d=d_L$

As argued in Section~\ref{sec5}, this case belongs to the additive scheme and is contained in the previous one.
From Eq.~(\ref{N5}) follows that $C_{\psi}(t,t_w,t_0)$, and therefore
also  $C_{st}(\tau)$, vanish identically in the quenches to $T=0$. Then we have 
$C(t,t_w,t_0) = C_{\sigma}(t,t_w,t_0)$ which, using Eqs.~(\ref{N1bis}) and~(\ref{N6}), is equivalent to Eq.~(\ref{SCR19}).
Conversely, $R_{st}(\tau)$ and $R_{ag}(t,t_w,t_0)$ are temperature independent and are still given by
Eqs.~(\ref{NN7}) and~(\ref{NN8}) with $d=2$ and $a=0$. Notice that the equilibrium FDT is satisfied also in the limit
$T \rightarrow 0$, where $C_{st}(\tau)$ vanishes, since from Eq.~(\ref{NN4}) follows
\be
\lim_{T \to 0} {1 \over T} {\partial \over \partial t_w} C_{st}(\tau) = R_{st}(\tau).
\label{NN88}
\ee

\end{enumerate}

We emphasize, here, that the explicit form~(\ref{N9bis}) of the exponent $a$ is a particular case
of Eq.~(\ref{4.2}) with $n=2, z=2, d_L=2$ and $d_U=4$. 
It should also to be remarked that $a$ is given by the same expression~(\ref{N9bis})
both in the quenches to $T_C$ and below $T_C$. This is a peculiarity of the large-$N$ model,
where $\eta=0$ and $z_c=z$.

\subsection{Effective temperature ${\cal T}(C)$} \label{sec6.2}

\begin{enumerate}

\item {\it Critical quench}  

From the explicit expressions for $f_C(x,y)$ and $f_R(x,y)$ in the critical quenches follows 
\begin{eqnarray}
T_{eff}(t,t_w,t_0)& = & F(x,y) \\ \nonumber
& = & dT_C  \left \{ (x+y)\int_0^1 dz z^{d/2-2} (x+1-2z+y)^{-1-d/2} \right .\\ \nonumber 
& - &  \left . {1 \over 2}\int_0^1 dz z^{d/2-2} (x+1-2z+y)^{-d/2} \right \}  (x-1+y)^{d/2}. 
\label{N10}
\end{eqnarray}
Evaluating numerically the right hand side, as $y \rightarrow 0$ the curve $F(x,y)$ approaches
(Fig.3) the limit function $F(x,0)$ rising from $F(1,0)=T_C$ to
$T_{\infty} =\lim_{x \to \infty}F(x,0) =T_C/X_{\infty}$, where $X_{\infty}= (d-2)/d$~\cite{Godreche-sph}.
Similarly, carrying out numerically the inversion of Eq.~(\ref{NN1}), the function  $x=f_C^{-1}(t_w^b C,y)$
shows (Fig.4) the approach to the singular behavior~(\ref{EFT4}) in the limit $y \rightarrow 0$.
The parametric plot in
Fig.5 displays the approach of $\widehat{T}(C,t_w)$ toward  ${\cal T}(C)$ of 
the form of Eq.~(\ref{I1}) as $t_w \rightarrow \infty$,
which can be very slow if $b$ is small. With the purpose of comparing further down with data for the
XY model, in Fig.6 we have also shown the parametric plot of the ZFC susceptibility.
This has been obtained by plotting $\chi(t,t_w)$ against $C(t,t_w)$ for fixed $t_w$
in two different cases: in panel (a), with $d=2.5$ corresponding to $b=0.25$, the approach to the asymptotic linear
plot~(\ref{XY3bis}) is evident, while in panel (b), with $d=2.06$ corresponding to the much smaller value $b=0.03$,
there is a much longer preasymptotic regime preceding the onset of the linear behavior.

\item {\it Quench below the critical  line}  

Using Eqs.~(\ref{NN7}) and~(\ref{NN8}) we may rewrite Eq.~(\ref{0003}) as
\be
T_{eff}(t,t_w,t_0) = Tx^{-d/4} + t_w^{d/2-1} F(x,y) (1-x^{-d/4})
\label{T1}
\ee
where
\be
F(x,y) = M^2(8 \pi)^{d/2} {d \over 4} {x^{d/4} \over (x^{d/4} -1)} \left ( 
{x-1+y \over x+1+y} \right )^{1+d/2}.
\label{T2}
\ee
In Fig.7 we have plotted $T_{eff}(t,t_w,t_0)$ against the exact $C(t,t_w,t_0)$ obtaining 
$\widehat{T}(C,t_w)$, which shows the approach toward the form~(\ref{I2}) of ${\cal T}(C)$
as $t_w$ grows.

\item {\it Quench to} $T=0$ {\it with} $d=d_L$

At $d=2$ and $T=0$, the second term in Eq.~(\ref{N1}) vanishes and $b_{\sigma}=0$, yielding
\be
C(x,y) = 2x^{1/2}(x+1+y)^{-1}
\label{N12}
\ee
while from Eqs.~(\ref{T1}) and~(\ref{T2})
\be
T_{eff}(t,t_w,t_0) = (4\pi)\left ( {x-1+y \over x+1+y} \right )^2.
\label{T3}
\ee
Letting $y \rightarrow 0$ and eliminating $x$ between the above two equations we obtain
the explicit non trivial expression
\be
{\cal T}(C) = (4\pi)\left ( {1-C^2 + \sqrt{1-C^2} \over 1+ \sqrt{1-C^2}}\right )^2  
\label{T4}
\ee
plotted in Fig.2c. We recall that the specific form of the function ${\cal T}(C)$ 
in the quench to $(d_L,T=0)$ is nonuniversal.

\end{enumerate}

\section{XY model} \label{sec7}

In the XY model the phase diagram of Fig.1 includes the KT line and, as anticipated in the Introduction,
in the $d=2$ case the behavior of ${\cal T}(C)$ for $0 < T \leq T_{KT}$ is expected to be given by
Eq.~(\ref{I1}). This is hard to detect
from existing data, since the small values of $b=\eta(T)/z$ at the temperatures used in~\cite{Mauro,Abriet2d}
would require to reach huge values of $t_w$ in order to observe the approach to~(\ref{I1}).
In Ref.~\cite{Mauro} data are presented for the ZFC susceptibility. According to
Eq.~(\ref{XY3}) asymptotically one should see the approach to the
linear parametric plot~(\ref{XY3bis}), as illustrated 
in Fig.6a for the large-$N$ model. The data in Fig.10 of Ref.~\cite{Mauro}
seem to be far from this behavior and, indeed, in~\cite{Mauro} these data have been regarded as  reminiscent of the non
trivial parametric plot in the $d=3$ Edwards-Anderson model. However, we believe that the non trivial 
pattern displayed by these data must be attributed to the large preasymptotic effect due to
the small value of $b=\eta(T)/z=0.03$. Indeed, the pattern displayed by $\chi(C,t_w)$ in Fig.10 of Ref.~\cite{Mauro}
is strikingly similar to the one in Fig.6b
for the critical quench in the large-$N$ model, despite the fact that in the large-$N$ model there is
no KT transition. The data in Fig.6b have been generated in order
to reproduce the same value of $b$ in the quench of the XY model considered in~\cite{Mauro}.

It would be interesting to have data
for the XY model quenched at $d_L$ and $T=0$, where, according to the general
argument of section~\ref{sec3}, $\chi(C)$ is expected to display the non trivial behavior discussed in
section~\ref{sec4}.

In the $d=3$ case in Ref.~\cite{Abriet3d}, the parametric plots of the FDR
\be
X(C)= \lim_{t_w \to \infty} {T \over \widehat{T}_{eff}(C,t_w)} = {T \over {\cal T}(C)}
\label{FDR}
\ee
are presented in the quenches to and below $T_C$. The data display the approach to the asymtotic parametric
forms
\begin{eqnarray} 
X(C) =   \left \{ \begin{array}{ll}
        X_{\infty}=T_C/T_{\infty} \qquad $for$ \qquad C =0\\
        1   \qquad $for$ \qquad   0 <C \leq 1
        \end{array}
        \right .
        \label{FDR1}
        \end{eqnarray} 
in the quench to $T_C$ and
\begin{eqnarray} 
X(C) =   \left \{ \begin{array}{ll}
        0 \qquad $for$ \qquad C < M^2\\
        1   \qquad $for$ \qquad M^2  \leq C \leq 1
        \end{array}
        \right .
        \label{FDR2}
        \end{eqnarray} 
in the quench below $T_C$, which correspond to Eqs.~(\ref{I1}) and~(\ref{I2}) for ${\cal T}(C)$.

Furthermore, in the $d=3$ model of
particular interest is the measurement of the exponent $a$ below $T_C$. The value
$a=1/2$ has been obtained both from the ZFC susceptibility~\cite{Generic} and 
from the direct mesurement of $R(t,t_w,t_0)$~\cite{Abriet3d}. This value gives additional and clearcut evidence
in favour Eq.~(\ref{4.2}), which, in fact, predicts $a=1/2$ when $d=3$ and the order 
parameter is non conserved $(z=2)$ and vectorial $(n=2,d_U=4)$. Conversely, the conjecture~\cite{Barrat} relating $a$ to
the density of defects, would predict  $a=1$, since in the XY model, with a vectorial order parameter, the density of defects
goes like $t^{-1}$~\cite{Bray,Blundell}.

\section{Conclusions} \label{sec8}

In this paper we have overviewed the behavior of the effective temperature in the slow relaxation
processes arising when systems, like a ferromagnet, with a simple pattern of ergodicity breaking
in the low temperature state,
are quenched from high temperature to or below $T_C$. On the basis of very general assumptions on
the scaling properties of the autocorrelation and response functions, we have derived the scaling 
behavior of the effective temperature, which allows to account in a unified way for the different
patterns displayed by ${\cal T}(C)$ in the different regions of the phase diagram. The primary
results are i)
in the critical quenches with $T_C>0$, as a consequence of $a=b>0$, ${\cal T}(C)$ 
displays always the form~(\ref{I1}), which is trivial in the sense that ${\cal T}(C) = T_C$ except for
the limiting value $T_{\infty}$ of Godr\`eche and Luck at $C=0$. In particular, this implies that 
also in the $d=2$ XY model quenched to $0<T \leq T_{KT}$ ${\cal T}(C)$, and the equivalent parametric representations
of the FDR or the ZFC susceptibility, must be trivial. The non trivial behavior reported in ref.~\cite{Mauro}, then, must be regarded as
preasymptotic. 
ii) The non trivial form of ${\cal T}(C)$ in the quenches to $(d_L,T=0)$ requires $a=0$. Since,
due to weak ergodicity breaking, this process cannot be regarded as the continuation to $T_C=0$
of the critical quenches, $a=0$ cannot be explained on the basis of Eq.~(\ref{207.1}). Rather,
the exponent $a$ in the quenches {\it below} $T_C$ must be expected to
have a dimensionality dependence such that $\lim_{d \to d_L}a=0$. The phenomenological formula~(\ref{4.2})
fits quite well within this framework, although a deeper theoretical understanding of the problem
is needed in order to put it on firmer grounds.

\vspace{5mm}

{\bf Acknowledgements}
 
We wish to thank Pasquale Calabrese, Silvio Franz and Mauro Sellitto for very useful discussions.
This work has been partially supported by MURST through PRIN-2002.

\begin{figure}
    \centering
   \rotatebox{0}{\resizebox{.65\textwidth}{!}{\includegraphics{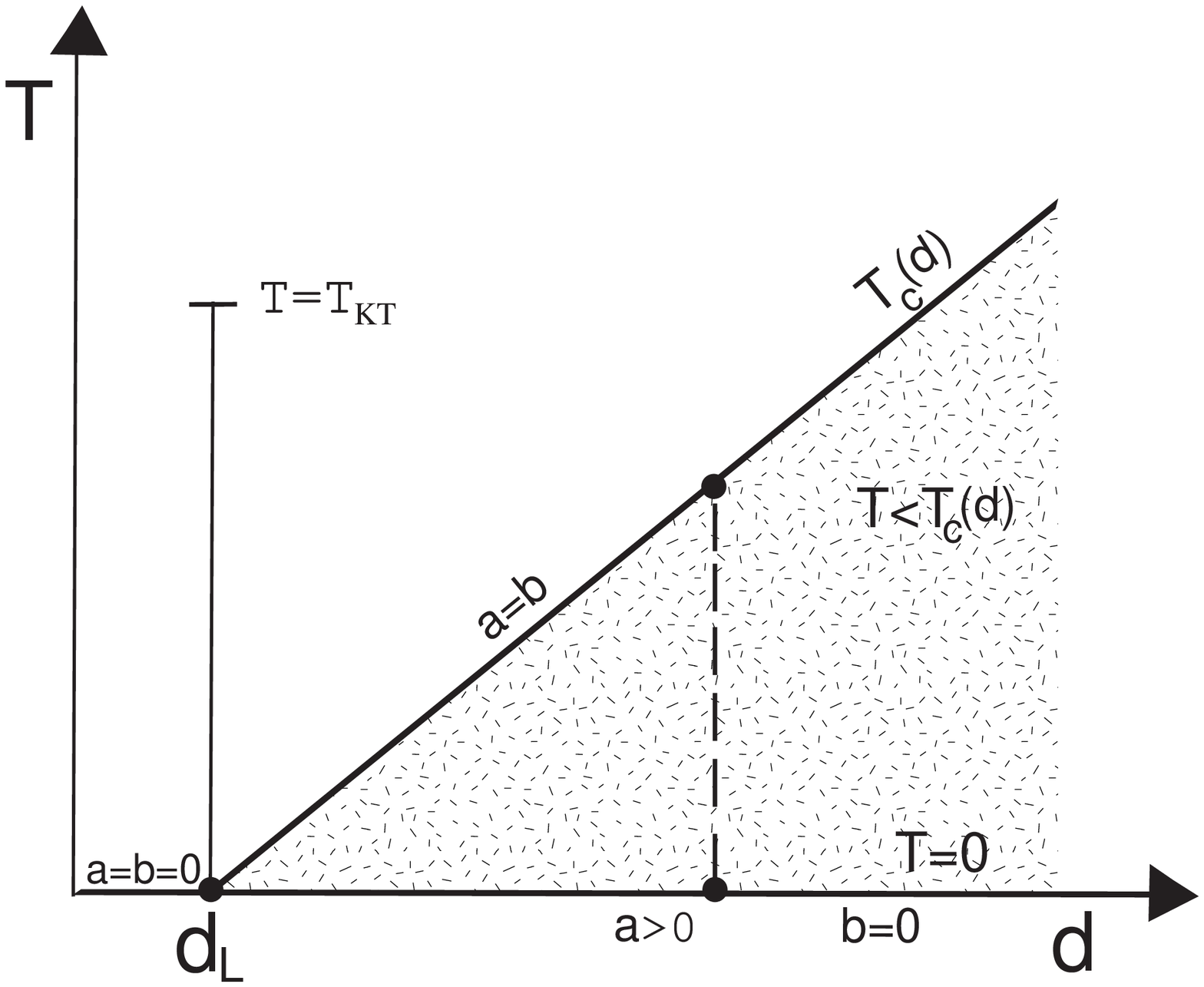}}}
    \caption{Schematic phase diagram in the $d-T$ plane with the critical line $T_C(d)$ and the
      Kosterlitz-Thouless line for the 2d XY model.}
    \label{fig1}
 \end{figure}

\newpage
\newpage

\begin{figure}
    \centering
   \rotatebox{0}{\resizebox{.65\textwidth}{!}{\includegraphics{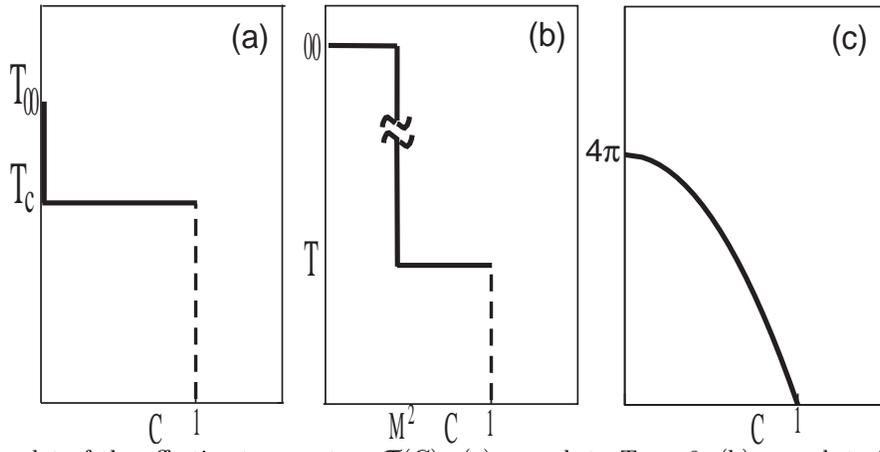}}}
    \caption{Parametric plot of the effective temperature ${\cal T}(C)$: (a) quench to $T_C > 0$, (b) quench to $T <T_C$
(c) quench to $(d_L,T=0)$. Panel (c) has been generated plotting the explicit form of ${\cal T}(C)$ given by
Eq.~(\ref{T4}) in the large-$N$ model.}
    \label{fig2}
 \end{figure}

\newpage

\newpage

\begin{figure}
    \centering
   \rotatebox{0}{\resizebox{.65\textwidth}{!}{\includegraphics{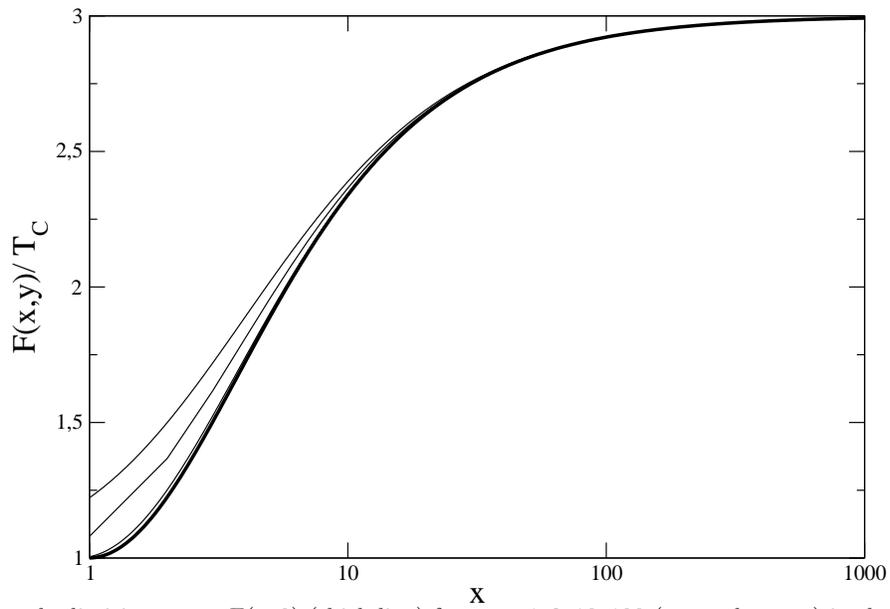}}}
    \caption{
Approach to the limiting curve $F(x,0)$ (thick line) for $t_w=1,2,10,100$ (top to bottom) in the 
critical quench of the 3d large-$N$ model.}
    \label{fig3}
 \end{figure}

\newpage

\newpage

\begin{figure}
    \centering
   \rotatebox{0}{\resizebox{.65\textwidth}{!}{\includegraphics{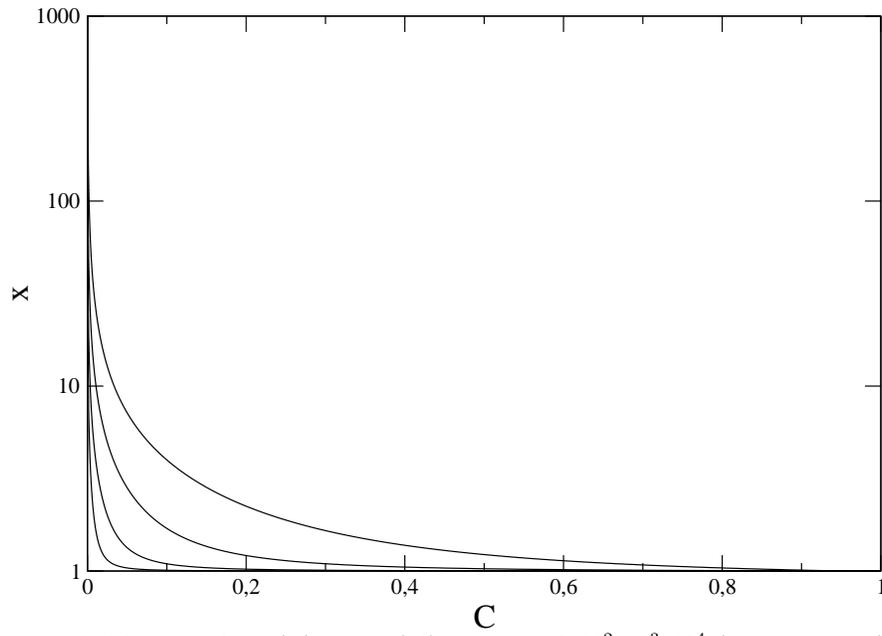}}}
    \caption{Approach to the limiting function $x(C)$ of Eq.~(\ref{EFT4}) for $t_w=10,10^2,10^3,10^4$ (top to bottom) in the 
critical quench of the 3d large-$N$ model.}
    \label{fig4}
 \end{figure}

\newpage

\newpage

\begin{figure}
    \centering
   \rotatebox{0}{\resizebox{.65\textwidth}{!}{\includegraphics{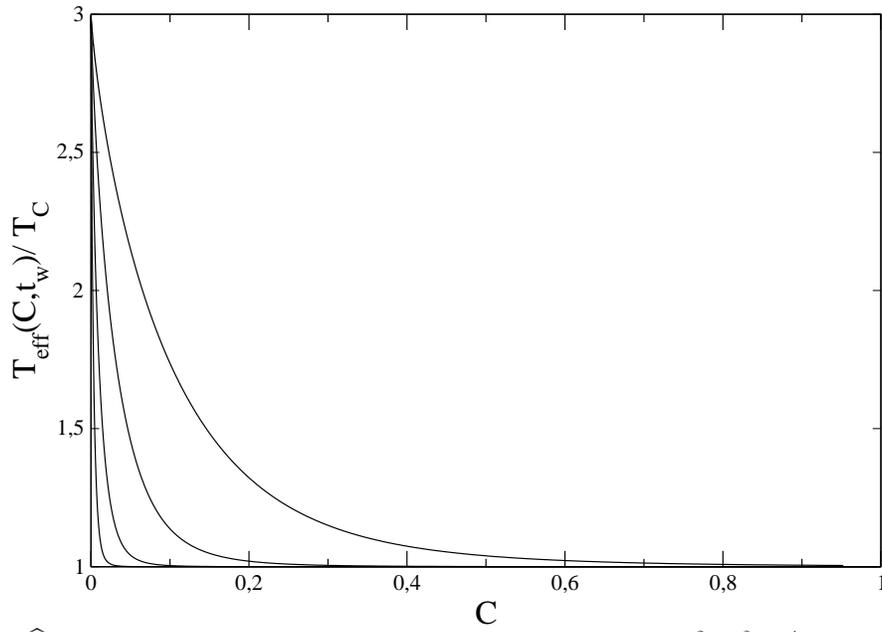}}}
    \caption{Approach of $\widehat{T}_{eff}(C,t_w)$ toward  ${\cal T}(C)$ of 
the form of Eq.~(\ref{I1}) for $t_w=10,10^2,10^3,10^4$ (top to bottom) in the critical quench of the 3d large-$N$ model.}
    \label{fig5}
 \end{figure}

\newpage

\begin{figure}
    \centering
   \rotatebox{0}{\resizebox{.65\textwidth}{!}{\includegraphics{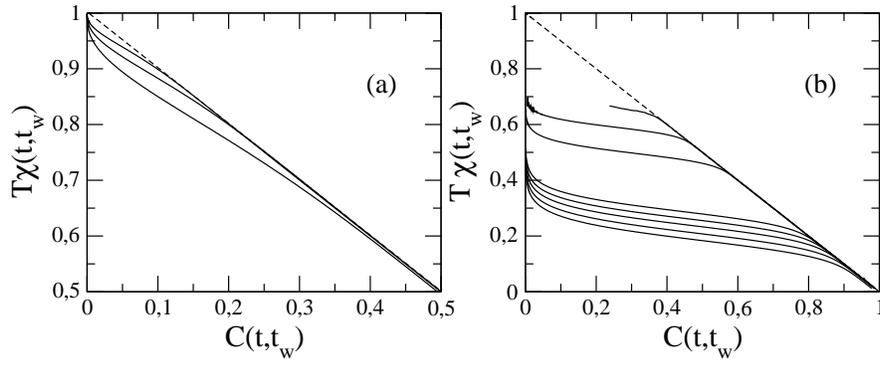}}}
    \caption{Parametric plot of the ZFC susceptibility in the critical quench of the large-$N$ model:
(a) $d=2.05$, $t_w=10^2,10^3,10^4$ (bottom to top), (b)  $d=2.06$, $t_w=10^2,3 \times 10^2,
10^3,3 \times 10^3,10^4,10^6,10^9,10^{12}$ (bottom to top).}
    \label{fig6}
 \end{figure}

\newpage

\begin{figure}
    \centering
   \rotatebox{0}{\resizebox{.65\textwidth}{!}{\includegraphics{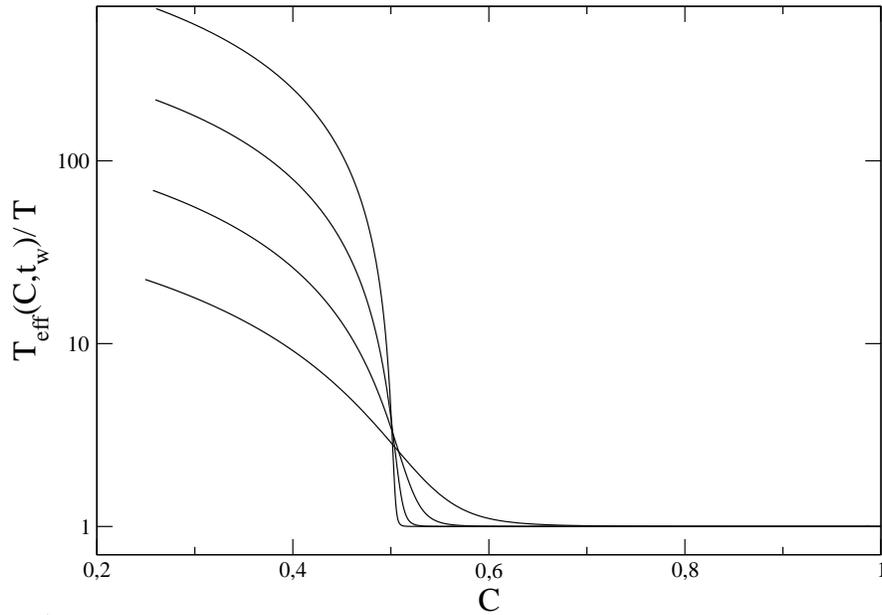}}}
    \caption{Approach of $\widehat{T}_{eff}(C,t_w)$ toward  ${\cal T}(C)$ of 
the form of Eq.~(\ref{I2}) for $t_w=10^2,10^3,10^4,10^5$ (bottom to top) in the quench to 
$T=T_c/2$ of the 3d large-$N$ model.}
    \label{fig8}
 \end{figure}

\newpage
\newpage

\newpage

\newpage


\begin{thebibliography}{99}
\bibitem{vector}
For systems with vector order parameter there is not, as of yet, enough information, analytical or 
numerical, to conjecture the general form of the response function.
\bibitem{note}
The equivalent formula for the FDR was derived in Ref.~\cite{Godreche}.
\end{thebibliography}
\end{document}